\documentclass{ceab}   

\usepackage{graphicx}   

\usepackage{ceabbib}     
\usepackage[T1]{fontenc}  
  \usepackage[scaled=0.94]{helvet}

  \usepackage{hyperref}
  \hypersetup{
      final=true,
      pageanchor=true,
      hyperfigures=false,
      colorlinks=true,
      breaklinks=true,
      linkcolor=blue,
      citecolor=blue,
      urlcolor=blue,
      pdfpagemode=UseNone,
      pdfpagelayout=OneColumn,
      pdfview=FitBH,
      pdftitle={Properties of a decaying sunspot},
      pdfauthor={Balthasar et al.},
      pdfsubject={Central European Astrophysical Bulletin}}

\begin{document}

\def\arcs{$^{\prime \prime}$\,}
\def\farcs{$^{\prime \prime}$\hspace{-0.15cm}.\,}
\def\aut{Balthasar et al.}
\def\tit{Properties of a decaying sunspot}

\title{Properties of a decaying sunspot}

\author{H. BALTHASAR$^{1}$, C. BECK$^2$, P. G\"OM\"ORY$^3$, K. MUGLACH$^4$, \\
K.~G. PUSCHMANN$^1$,  T. SHIMIZU$^5$ and M. VERMA$^{1}$
\vspace{2mm}\\
\it $^1$Leibniz-Institut f\"ur Astrophysik Potsdam, An der Sternwarte 16, \\
    D--14482 Potsdam, Germany\\ 
\it $^2$Instituto de Astrof{\'\i}sica de Canarias, Via Lact\'ea, \\
    E--38205 La Laguna, Tenerife, Spain\\
\it $^3$Astronomical Institute of the Slovak Academy of Science, \\
    SK--05960 Tatransk\'a Lomnica, Slovakia\\ 
\it $^4$Artep, Inc. at Goddard Space Flight Center  Washington DC, USA \\
\it $^5$Institute of Space and Astronautical Science, Japan Aerospace Exploration \\
        Agency, 3-1-1 Yoshinodai, Chuo-ku, Sagamihara, 252-5210 Kanagawa, Japan
}

\maketitle

\begin{abstract}
A small decaying sunspot was observed with the Vacuum Tower
Telescope (VTT) on Tenerife and the Japanese \textit{Hinode}
satellite. We obtained full Stokes scans in several
wavelengths covering different heights in the solar
atmosphere. Imaging time series from \textit{Hinode} and the Solar 
Dynamics Observatory (SDO) complete our data sets.
The spot is surrounded by a moat flow, which persists also
on that side of the spot where the penumbra already had disappeared.
Close to the spot, we find a chromospheric location with 
downflows of more than 10\,km\,s$^{-1}$ without photospheric counterpart.
The height dependence of the vertical component of the magnetic field strength 
is determined in two different ways that yielded different results 
in previous investigations. Such a difference still exists in our 
present data, but it is not as pronounced as in the past.
\end{abstract}


\section{Introduction}

Sunspots have been an important topic in solar physics
for decades, but still the detailed processes causing their 
fine structure are not well understood. Nevertheless, 
considerable progress in numerical modeling of sunspots
has recently been achieved by \citet{rempel11}.
One special issue is the height dependence of the magnetic field
in photospheric layers.
There is a long-lasting discrepancy between different methods to
determine the vertical derivative of the vertical component of
the magnetic field $B_{\rm z}$. Estimates depending on div\,${\bf B}$ = 0 
yield values below 1\,G\,km$^{-1}$ \citep{hagyardetal83, hofmanrendtel89, 
balthasar06, balthasargoemoery08}.
On the other hand, gradients of the magnetic field 
determined from two different spectral lines yield
values between 1.5 and 3\,G\,km$^{-1}$, not only for $B_{\rm z}$,
but also for the total magnetic field strength
\citep{wittmann74, balthasarschmidt93, balthasargoemoery08}. 
While the first method requires a high
spatial resolution for an accurate determination of the needed
horizontal derivatives, the latter method depends
on a proper knowledge of the height difference between the layers
in which the spectral lines predominantly form.
   \begin{figure}[!t]
  \begin{center}
\includegraphics[width=125mm]{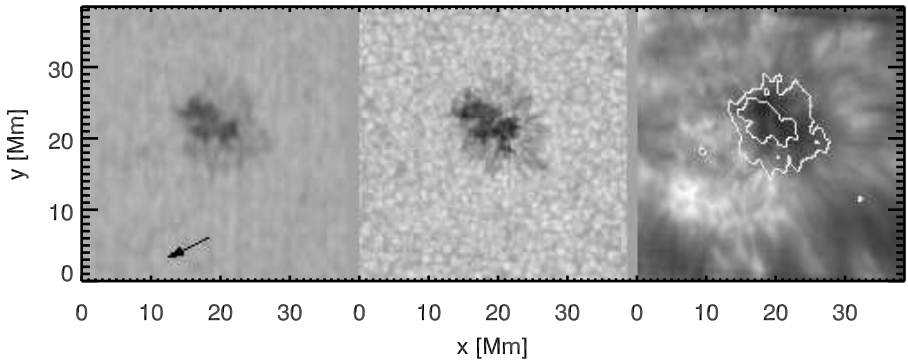}
    \end{center}
  \caption{Intensity images created from TIP (left) and \textit{Hinode} 
      (middle) scans and the line core image of the Ca\,{\sc ii}\,854.2\,nm 
      line. White contours show the umbral and penumbral boundary of the TIP
      data. The arrow points towards disk center.}
\label{fig_int}
  \end{figure} 

Sunspots are usually surrounded by a moat flow, an outward flow
first detected by \citet{sheely72}. At the outer boundary,
the moat is separated from neighboring supergranules by the magnetic
network. This flow is sometimes considered as an extension of the Evershed
flow.
\citet{martinezpilletetal09} obtained evidence for the continuation of
the Evershed flow beyond the sunspot boundary even connected with
supersonic velocities.
\citet{vargasetal07} found that there is no moat flow where
the umbra reaches the outer spot boundary without a penumbra in between.

\section{Observations and data processing}

The small decaying sunspot in active region NOAA 11277
was observed on September 2, 2011 with the Vacuum Tower Telescope
(VTT) on Tenerife and the Japanese \textit{Hinode} satellite
\citep{kosugietal07}.
In addition, we use data from the Helioseismic and Magnetic Imager \citep[HMI,][]{shouetal12a, shouetal12b, couvidatetal12}
on board the Solar Dynamics Observatory (SDO).
The spot was located at 19$^\circ$N and 30$^\circ$W 
($\vartheta$=31$^\circ$).

   \begin{figure}[!ht]
  \begin{center}
\includegraphics[width=125mm]{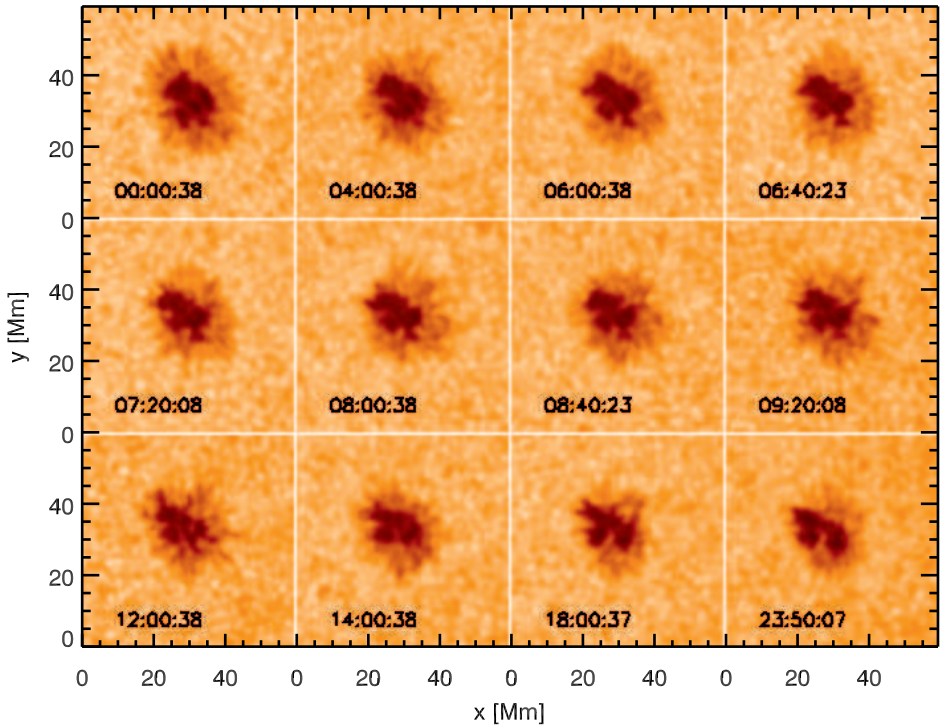}
    \end{center}
  \caption{
       Intensity images from HMI/SDO showing the evolution of the spot on 
September 2, 2011.
}
\label{fig_sdo}
  \end{figure} 

At the VTT, we used the Tenerife Infrared Polarimeter 2 \citep[TIP,][]{colladosetal07} 
attached to the spectrograph to scan a map of the sunspot. We selected the spectral lines Fe\,{\sc i}\,1078.3\,nm and 
Si\,{\sc i}\,1078.6\,nm, which both split into Zeeman triplets with
an effective Land\'e factor of 1.5.
The scan consists of 150 steps with a 0\farcs36 spacing.
Four independent exposures
with different states of the ferro-electric liquid crystals are required to 
obtain the full Stokes-vector, and we accumulated eight cycles. A single 
exposure took 250\,ms, thus, about 10\,s were needed per scan step.
The scan was started at 08:09\,UT and lasted 26\,min.
In addition to TIP, we mounted a CCD-camera centered on the line 
Ca\,{\sc ii}\,854.2\,nm \citep[see][]{becketal12}. Because this detector could not be 
synchronized with the polarimeter and has a low 
sensitivity at this wavelength, we integrated over 7.5\,s.
The image was stabilized by the Kiepenheuer Adaptive Optics System \citep[KAOS,][]{vdletal03}.
A continuum image recombined from the TIP scans and a core image 
from the ionized calcium line are given in Fig. \ref{fig_int} 
together with the corresponding area covered by the \textit{Hinode} scans.

The spot was in its decaying phase as shown in Fig~\ref{fig_sdo},
which displays selected HMI intensity images.
The total series consists of 144 measurements taken every 10\,min
during September 2. At 00:00\,UT, the umbra was still
completely surrounded by a penumbra, while the umbra was already irregular
but still continuous. During the time of the spectropolarimetric 
observations, the penumbra had almost disappeared on the eastern 
side. Later, the umbra split into two parts.

 \begin{figure}[!ht]
  \begin{center}
\includegraphics[height=135mm]{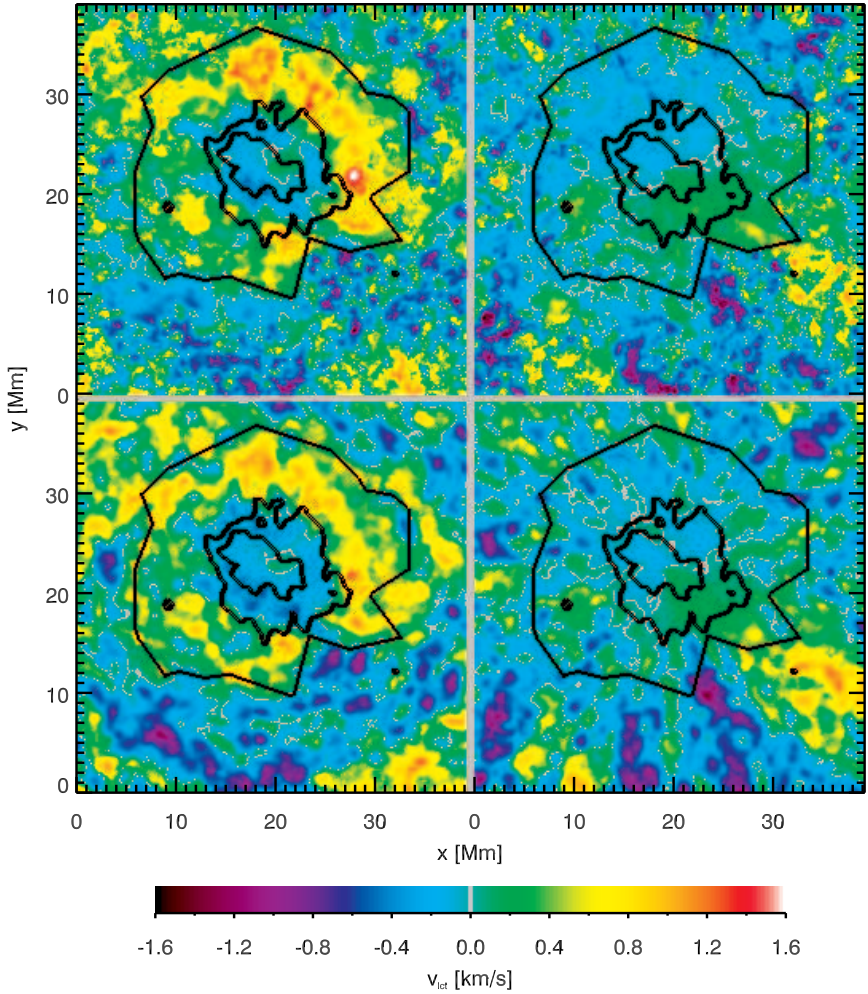}
 \end{center}
  \caption{Horizontal velocities derived from LCT. The left column 
      displays the 
      radial component, the right column exhibits the tangential 
      component. The upper panels show results from the Ca\,{\sc ii}\,H filtergrams
      and the lower ones those from G-band. White contours show the 
      boundary of the umbra, black contours the outer edge of the penumbra 
      and the moat. Positive values indicate radial velocities away from 
      the spot or counterclockwise tangential motions.
    }
\label{fig_lct}
  \end{figure} 

With the \textit{Hinode} satellite we used the spectropolarimeter 
\citep[HSP,][]{ichimotoetal08} to take another map of the full Stokes-vector in the lines 
Fe\,{\sc i}\,630.15\,nm and Fe\,{\sc i}\,630.25\,nm. This map 
was started at 07:34 UT covering an area of 123\arcs $\times$ 123\arcs
with 400 steps and a spacing of 0\farcs3. In addition, we took filtergrams in Ca\,{\sc ii}\,H
and G-band every 32\,s with the Broad-band Filter Imager (BFI) 
of the Solar Optical Telescope \citep[SOT,][]{tsunetaetal08}. These 
filtergrams cover an area of 111\arcs
$\times$ 111\arcs{} with a pixel size of 0\farcs11 $\times $
0\farcs11. Such images were obtained over a period of three hours.

Strength and orientation of the magnetic field, and Doppler velocities 
from the photospheric lines were derived with the code Stokes Inversion 
based on Response functions \citep[SIR,][]{ruizcobodeltoroiniesta92}.
We performed the inversions separately for the four spectral lines
and considered the magnetic field and velocity to be height independent 
at least for the height range where the line profiles are affected by 
these quantities. Only the temperature was height dependent with three nodes.
In preparation of the inversions, we subtracted 8\% undispersed straylight 
from the TIP data and kept the amount of dispersed straylight fixed at 5\% 
\citep[typical VTT values, see][]{becketal11}.
The magnetic azimuth ambiguity was resolved  assuming that there 
is an azimuth center in the spot. This approach was sufficient for the TIP data.
For the spatially more extended maps obtained with \textit{Hinode}, we applied in 
addition the minimum energy code of \citet{lekaetal09}. The coordinate system of the 
magnetic field was then rotated to the local reference frame.
Finally, we corrected the geometrical foreshortening with the 
method described by \citet{vermaetal12}.

 \section{Results}

\subsection{Velocity fields}

The time series of \textit{Hinode} filtergrams in Ca\,{\sc ii}\,H and G-band were the
base for a local correlation technique analysis (LCT) using the 
tools of \citet{vermadenker11}. In Fig~\ref{fig_lct}, we separate the horizontal 
velocities into a radial and a tangential component with respect 
to the center of the spot, as presented before by \citet{balthasarmuglach10}. 

The moat flow is well visible around the spot, and it reaches a distance 
of another spot radius. This extension is less than the one investigated 
by \citet{balthasarmuglach10}, but still within the range given by 
\citet{sobotkaroudier07}. Velocities decrease from about 1000 m s$^{-1}$
at the penumbral boundary to about 200 m s$^{-1}$ at the outer edge of 
the moat. The moat flow is clearly detectable also on the side where 
the penumbra disappeared, but velocities close to the umbra are much 
smaller than on the opposite side of the spot.
The tangential component is small inside the moat.
Beyond the moat, the signature of the supergranulation 
becomes visible. The nearest neighboring supergranules cause a 
ring of velocities towards the spot, while an alternating pattern 
can be seen in the tangential component.
 \begin{figure}[!ht]
  \begin{center}
\includegraphics[width=125mm]{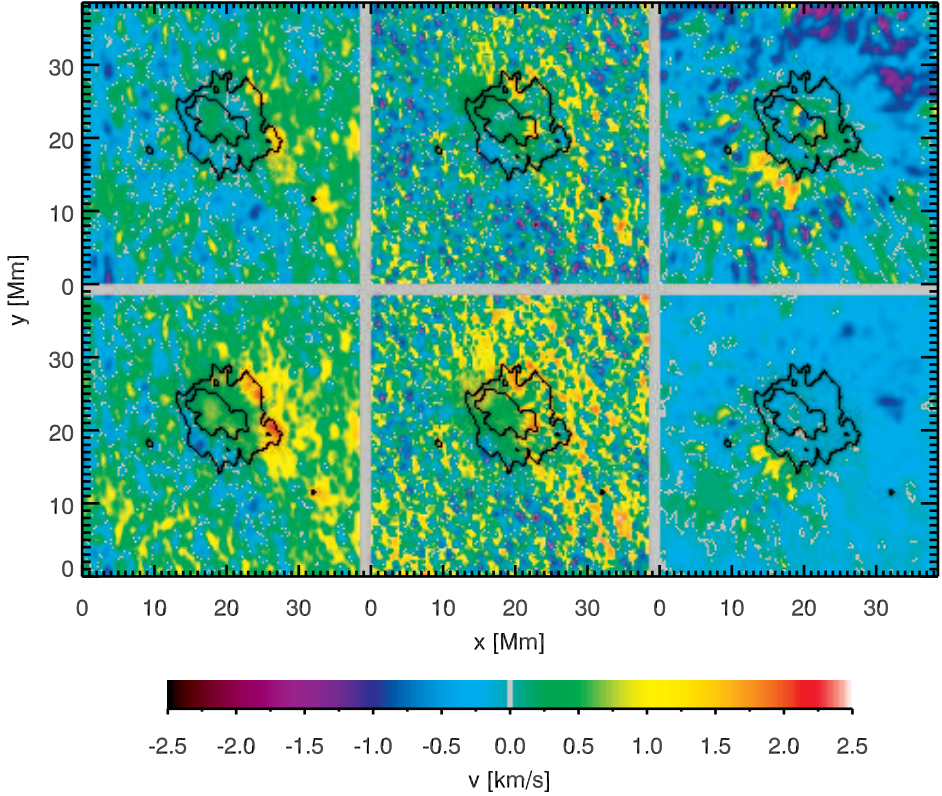}
 \end{center}
  \caption{Doppler velocities from TIP (left, Si: upper panel, Fe: lower panel), 
      from HSP (middle, Fe\,630.15 top, Fe\,630.25 bottom)
      and the infrared Ca line (right, line minimum top, bisector bottom).
      The chromospheric Ca\,{\sc ii} IR velocities have been reduced by a factor of 
      five to fit onto the same scale bar. Negative (blue) values correspond to 
      blueshifts.
    }
\label{fig_doppler}
  \end{figure} 

From the spectropolarimetric data, we also obtained Doppler 
velocities during the inversion process.
To the core of the Ca\,{\sc ii}\,854.2\,nm line, we applied a 
polynomial fit of fourth degree and calculated its minimum position,
which gives us the velocity. In addition, we derived the bisector 
velocity at 0.275 of the continuum intensity.
The line-of-sight velocities  are shown in Fig~\ref{fig_doppler}.
The Evershed effect is rather confined to deep photospheric layers.
We see positive velocities outside the spot on the limb side, and a dominance 
of negative values on the center side. These velocities and their extensions 
are in agreement with the moat flow found by LCT.
The most pronounced feature in the chromosphere is a downflow of more than 
10\,km\,s$^{-1}$ next to the spot on the center side. This downflow has no 
obvious counter part in photospheric layers. On the limb side, we see
negative velocities, which are probably related to the inverse Evershed effect 
in the chromosphere.

\subsection{The magnetic field}

\begin{figure}[!ht]
  \begin{center}
\includegraphics[width=110mm]{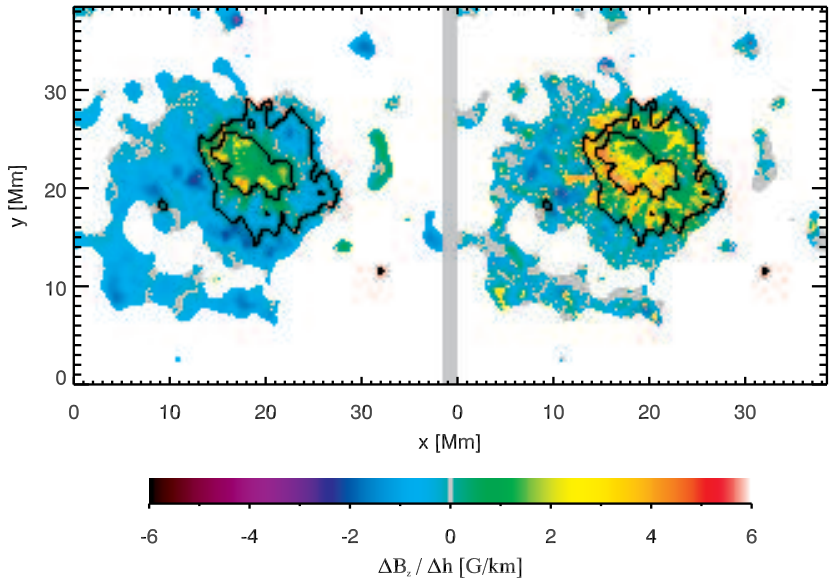}
    \end{center}
  \caption{Gradients of the vertical magnetic field derived from the difference method
      from TIP (left) and \textit{Hinode} (right) data.
    }
\label{fig_delta}
  \end{figure} 

The height dependence of $B_{\rm z}$
was derived in two different ways. The first method was to take the 
difference of $B_{\rm z}$ from two spectral lines divided by their height 
difference (difference method).
To estimate the height difference, we used the temperature maps obtained
for log\,$\tau$ = 0  and interpolated between the formation heights of 
the lines in a  quiet-Sun atmosphere \citep[Harvard Smithsonian 
Reference Atmosphere,][]{gingerichetal71} and in the umbral model M4 of \citet{kollatschnyetal80}.
For the two iron lines from HSP, the height differences obtained this way
are close to the 64\,km determined by \citet{faurobertetal09}
in the quiet Sun.
This method was performed for the two TIP lines and the two
HSP lines. The results for $\Delta B_z / \Delta z$ are displayed 
in Fig.~\ref{fig_delta}.

The spot had a negative polarity, therefore $B_{\rm z}$
is negative, and a decrease of $|B_{\rm z}|$ 
with height appears as a positive number. In the umbra,
we encounter a decrease of $|B_{\rm z}|$ by about 1\,G\,km$^{-1}$
for the TIP data. Only close to the penumbral boundary, some
locations exhibit up to 5\,G\,km$^{-1}$. The decrease is much stronger
for the HSP data. Here, we find values around 3\,G\,km$^{-1}$
and more locations with about 5\,G\,km$^{-1}$.
In the penumbra, we find an increase of $|B_{\rm z}|$ for the TIP data,
while the HSP data still exhibit a decrease. A possible explanation can 
be that Fe\,{\sc i} 1078.3\,nm is formed in deep layers, and the 
penumbral magnetic field might reside already above these layers.

The second method (derivative method) made use of the 
condition:

$$ \rm{div}{\bf B} = \frac{\partial B_x}{\partial x} + 
                  \frac{\partial B_y}{\partial y} + 
                  \frac{\partial B_z}{\partial z} = 0 $$

 \begin{figure}[!ht]
  \begin{center}
\includegraphics[width=110mm]{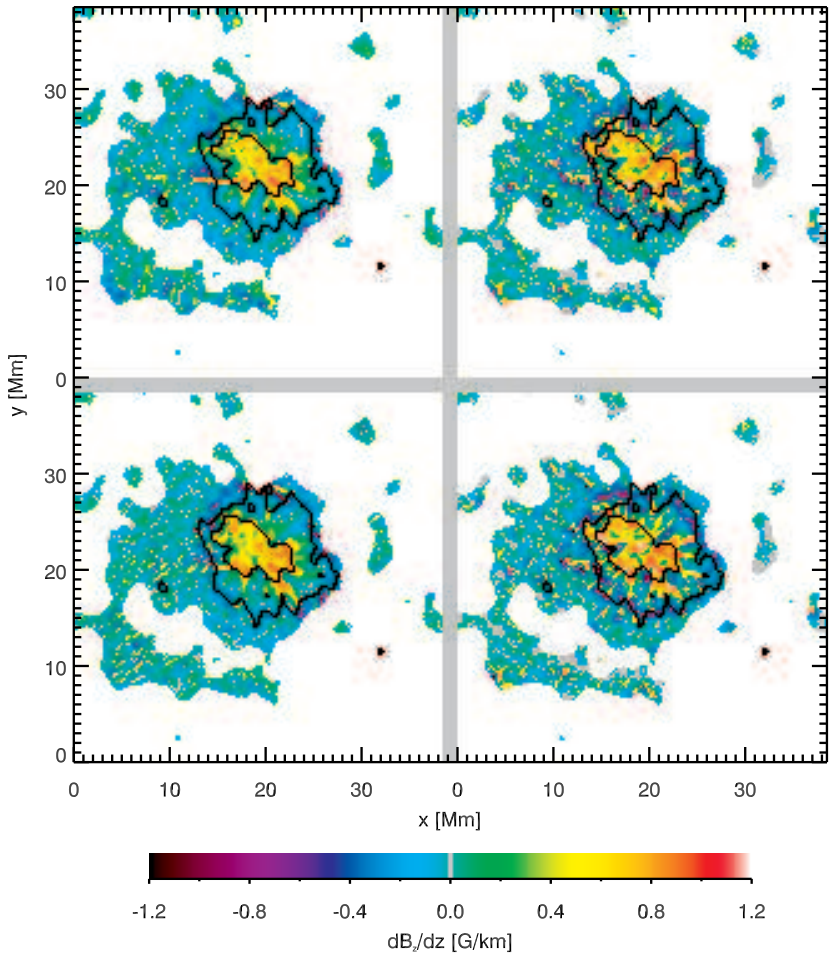}
 \end{center}
  \caption{Vertical derivatives of the vertical component of the
      magnetic field strength derived from ${\rm div} {\bf B} = 0$ for 
      Si\,{\sc i}\,1078.6\,nm (upper left), Fe\,{\sc i}\,1078.3\,nm
      (lower left), Fe\,{\sc i}\,630.15\,nm (upper right), and
      Fe\,{\sc i}\,630.25\,nm (lower right).
    }
\label{fig_dbzdz}
  \end{figure} 

This equation states that the vertical derivative must be compensated 
by the horizontal ones. The horizontal derivatives are determined from
the difference of the values in the neighboring pixels \citep{balthasar06}.
Such vertical derivatives were determined for all four
lines and are shown in Fig.~\ref{fig_dbzdz}. 
With this method, we find maximum values of about 1\,G\,km$^{-1}$ for
${\partial B_z}/{\partial z}$ from all four lines in 
the umbra, but also in some locations in the penumbra. For the outer 
penumbra, we encounter the opposite sign for this derivative.
These results are in a good agreement with difference results 
obtained from the TIP data, while there are discrepancies for the
HSP data. Remarkable is the fine structure seen in the HSP results,
indicating the importance of high spatial resolution to obtain
proper results from this method.

\section{Discussion and Conclusions}

The problem of the discrepancy between different methods to determine the 
height dependence of the magnetic field still exists, but there are indications
that it becomes smaller with higher spatial resolution. Nevertheless,
the difference method yields high values for the HSP data, but especially
for these lines, the height differences are reliable.
It would be much easier to argue that the formation layers of the TIP 
lines are not accurate enough, because the two lines are formed close 
to each other in the umbra, and it could be that they are closer together 
than estimated, resulting in a larger height gradient.
A correct determination of the height dependence of the magnetic field might be 
provided in the near future by extending the geometric height scale algorithm of 
\citet{puschmannetal10} to an entire sunspot. So far, this method has been applied only to a small region of a sunspot.
The spot is surrounded by an outward moat flow, as it is clearly visible from 
the LCT analysis. Photospheric Doppler shifts show opposite signs 
on both sides of the spot,
and they can be interpreted as an outward flow, too.
The extension of the moat is about one radius of the spot, in agreement with previous investigations, \citep[e.g.][]{sobotkaroudier07}. 
We still detect the moat flow on the side where the penumbra already disappeared, but more pronounced in the outer 
moat. This finding agrees with \citet{dengetal07}, but disagrees with the results 
of \citet{vargasetal07}. Probably one has to distinguish between cases where a penumbra did 
not form because of the presence of another spot or pore of the same polarity 
\citep{kuenzel69} and cases where the penumbra disappears because the spot is decaying.
In the chromosphere, the Doppler shifts have the opposite direction and exhibit 
the inverse Evershed effect. Remarkable is a downflow of more than 10\,km\,s$^{-1}$
close to the spot which has no photospheric counterpart.

With the next generation of solar telescopes that are going into operation now,
the New Solar Telescope at the Big Bear Solar Observatory \citep{caoetal10}
and GREGOR solar telescope at Tenerife \citep{schmidtetal12} a big step towards better resolution will be
done, and that will shed more light on the problems investigated in 
this contribution.

\section*{Acknowledgments} 

The Vacuum Tower Telescope in Tenerife is operated by the
Kiepenheuer-Institut f\"ur Sonnenphysik (Germany) at the 
Spanish Observatorio del Teide of the Instituto de
Astrof\'\i sica de Canarias.
\textit{Hinode} is a Japanese mission developed and launched by ISAS/JAXA, 
with NAOJ as domestic partner and NASA and STFC (UK) as international 
partners. It is operated by these agencies in cooperation with ESA and NSC 
(Norway).
The data have been used by courtesy of NASA/SDO and the HMI science team.
M.V. thanks the DAAD for its support in form of a PhD scholarship. 
P.G. acknowledges support by VEGA grant 2/0108/12.

\section*{References}
\begin{itemize}
\small
\itemsep -2pt
\itemindent -20pt
\bibitem [{Balthasar(2006)}]{balthasar06} 
 Balthasar, H.: 2006, {\it \aap} {\bf 449}, 1169.
\bibitem [{Balthasar \& G\"om\"ory(2008)}]{balthasargoemoery08} 
 Balthasar, H. and G\"om\"ory, P.: 2008, {\it \aap} {\bf 488}, 1085.
\bibitem [{Balthasar \& Muglach(2010)}]{balthasarmuglach10} 
 Balthasar, H. and Muglach, K.: 2010, {\it \aap} {\bf 511}, A67.
\bibitem [Balthasar \& Schmidt(1993)]{balthasarschmidt93} 
 Balthasar, H. and Schmidt, W.: 1993, {\it \aap} {\bf 279}, 243.
\bibitem [{Beck et al.(2011)}]{becketal11} 
  Beck, C., Rezaei, R., and Fabbian, D.: 2011, {\it \aap} {\bf 535}, A129.
\bibitem [{Beck et al.(2012)}]{becketal12} 
  Beck, C., Rezaei, R., and Puschmann, K.~G.: 2012, {\it \aap} {\bf 544}, A46.
\bibitem [{Cao et al.(2010)}]{caoetal10} Cao, W., Gorceix, N., Coulter, R. et al.: 2010, {\it Astron. Nachr.} 
       {\bf 331}, 636.
\bibitem [{Collados et al.(2007)}]{colladosetal07} 
  Collados, M., Lagg, A., D{\'\i}az~Garc{\'\i}a, J.J. et al.: 2007,in: P. Heinzel, I. Dorotovi{\v c}, R.J. Rutten (eds.),  
             {\it The physics of chromospheric plasmas ASPCS} {\bf 368}, 611.
\bibitem [{Couvidat et al.(2012)}]{couvidatetal12} 
   Couvidat,S., Schou, J., Shine, R.~A. et al.: 2012, {\it Sol. Phys.} {\bf 275}, 285.
\bibitem [{Deng et al.(2007)}]{dengetal07} 
  Deng, N., Choudhary, D.~P., Tritschler, A. et al.: 2007,{\it \apj} {\bf 671}, 1013.
\bibitem [{Faurobert et al.(2009)}]{faurobertetal09} 
  Faurobert, M., Aime, C., P\'erini, C. et al.: 2009, {\it \aap} {\bf 507}, L29.
\bibitem [{Gingerich et al.(1971)}]{gingerichetal71} Gingerich, O., Noyes, R.~W., Kalkofen, W., and Cuny, Y.: 1971, {\it Sol. Phys.} 
       {\bf 18}, 347.
\bibitem [{Hagyard et al.(1983)}]{hagyardetal83} 
        Hagyard, M.~J., Teuber, D., West, E.~A. et al.: 1983, {\it Sol. Phys.} {\bf 84}, 13.
\bibitem [{Hofmann \& Rendtel(1989)}]{hofmanrendtel89} 
      Hofmann, A. and Rendtel, J.: 1989, {\it Astron. Nachr.} {\bf 310}, 61. 
\bibitem [{Ichimoto et al.(2008)}]{ichimotoetal08} 
   Ichimoto, K. Lites, B.~W., Elmore, D. et al.:  2008, {\it Sol. Phys.} 
        {\bf249}, 233.
\bibitem [{Kollatschny et al.(1980)}]{kollatschnyetal80} 
  Kollatschny, W., Stellmacher, G., Wiehr, E., and Falipou, M.~A.: 1980,
     {\it \aap} {\bf 86}, 245.
\bibitem [{Kosugi et al.(2007)}]{kosugietal07} 
  Kosugi, T., Matsuzaki,K., Sakao, T. et al.: 2007, {\it Sol. Phys.} {\bf243}, 3. 
\bibitem [{K\"unzel(1969)}]{kuenzel69} K\"unzel, H.: 1969, {\it Astron. Nachr.} {\bf 291}, 265.
\bibitem [{Leka et al.(2009)}]{lekaetal09} 
  Leka, K.~D., Barnes, G., Crouch, A.~D. et al.: 2009, {\it Sol. Phys.} {\bf 260}, 83.
\bibitem [{Mart{\'\i}nez Pillet et al.(2009)}]{martinezpilletetal09} 
   Mart{\'\i}nez Pillet, V., Katsukawa, Y., Puschmann, K.~G., and Ruiz~Cobo, B.: 2009, {\it \apj} {\bf 701}, L79.
\bibitem [{Puschmann et al.(2010)}]{puschmannetal10} 
  Puschmann, K.~G., Ruiz Cobo, B., and Mart{\'\i}nez Pillet, V.: 2010, {\it \apj}
         {\bf 720}, 1417.
\bibitem [{Rempel(2011)}]{rempel11} 
         Rempel, M.: 2011, {\it \apj} {\bf 729}, 5.
\bibitem [{Ruiz~Cobo \& del~Toro~Iniesta(1992)}]{ruizcobodeltoroiniesta92} 
  Ruiz~Cobo, B. and del~Toro~Iniesta, J.C.: 1992, {\it \apj} {\bf 398}, 375.
\bibitem [Schmidt et al.(2012)]{schmidtetal12} 
  Schmidt,W., von der L\"uhe, O., Volkmer, R. et al.: 2012, {\it Astron. Nachr.} {\bf 333}, 796.
\bibitem [{Shou et al.(2012b)}]{shouetal12b}
  Schou, J., Borrero, J.~M., Norton, A.~A. et al.: 2012b, {\it Sol. Phys.} {\bf 275}, 327.
\bibitem [{Shou et al.(2012a)}]{shouetal12a}  
  Schou, J., Scherrer, P.~H., Bush, R.~I. et al.: 2012a, {\it Sol. Phys.} {\bf 275}, 229.
\bibitem [{Sheely(1972)}]{sheely72} 
   Sheeley, N.~R.: 1972, {\it Sol. Phys.} {\bf 25}, 98.
\bibitem [{Sobotka \& Roudier(2007)}]{sobotkaroudier07} 
   Sobotka, M. and Roudier, T.: 2007, {\it \aap} {\bf 472}, 277.
\bibitem [{Tsuneta et al.(2008)}]{tsunetaetal08} Tsuneta, S., Ichimoto, K., Katsukawa, Y. et al.: 2008, {\it Sol. Phys.}
       {\bf249}, 167.
\bibitem [{Vargas Dom{\'\i}nguez et al.(2007)}]{vargasetal07} 
   Vargas Dom{\'\i}nguez, S., Bonet, J.~A., and Mart{\'\i}nez Pillet, V. et al.: 2007, {\it \apj} {\bf 600}, L165.
\bibitem [{Verma \& Denker(2011)}]{vermadenker11} 
  Verma, M. and Denker, C.: 2011, {\it \aap} {\bf 529}, A153.
\bibitem [{Verma et al.(2012)}]{vermaetal12} 
  Verma, M., Balthasar, H., Deng., N. et al.: 2012, {\it \aap} {\bf 538}, A109.
\bibitem [{von der L\"uhe et al.(2003)}]{vdletal03} 
  von der L\"uhe, O., Soltau, D., Berkefeld, T., and Schelenz, T.: 2003,
   {\it Proc. SPIE} {\bf 4853}, 187.
\bibitem [{Wittmann(1974)}]{wittmann74} Wittmann, A.: 1974, {\it Sol. Phys.} {\bf 36}, 29.

\end{itemize}

\end{document}